# A lot of talk and a badge: An exploratory analysis of personal achievements in GitHub

Fabio Calefato [*], Luigi Quaranta, Filippo Lanubile

*University of Bari, Dipartimento di Informatica, Via E. Orabona, 4, Bari, 70125, Italy*

A B S T R A C T

**Context:** GITHUB has introduced a new gamification element through personal achievements, whereby badges are unlocked and displayed on developers' personal profile pages in recognition of their development activities.
**Objective:** In this paper, we present an exploratory analysis using mixed methods to study the diffusion of personal badges in GITHUB, in addition to the effects and reactions to their introduction.
**Method:** First, we conduct an observational study by mining longitudinal data from more than 6,000 developers and performed correlation and regression analysis. Then, we conduct a survey and analyze over 300 GITHUB community discussions on the topic of personal badges to gauge how the community responded to the introduction of the new feature.
**Results:** We find that most of the developers sampled own at least a badge, but we also observe an increasing number of users who choose to keep their profile private and opt out of displaying badges. Additionally, badges are generally poorly correlated with developers' skills and dispositions such as timeliness and desire to collaborate. We also find that, except for the `Starstruck` badge (reflecting the number of followers), their introduction does not have an effect. Finally, the reaction of the community has been in general mixed, as developers find them appealing in principle but without a clear purpose and hardly reflecting their abilities in the current form.
**Conclusions:** We provide recommendations to the designers of the GITHUB platform on how to improve the current implementation of personal badges as both a gamification mechanism and as sources of reliable cues for assessing the abilities of developers.

## 1. Introduction

Gamification refers to the practice of incorporating elements and mechanics of game design into non-game contexts [1,2] to increase user engagement and motivation, as well as drive desired behaviors by creating a sense of play and/or competition [3,4]. Previous research has found evidence that gamification mechanisms are in part responsible for the success of websites such as STACK OVERFLOW (SO) [5]. Gamification, in fact, has always been a staple of the technical Q&A site whose members gain badges, medals, and privileges by being good community citizens (e.g., answering questions, editing poorly written posts, etc.). Another popular source of user-generated content in the Software Engineering domain is GITHUB, which, in contrast, has always incorporated several social features (e.g., developers have their personal profiles and can follow each other, repositories can be starred), but only implemented a few and simple gamification elements on its website, such as code streak counters, which were removed from the contribution graph in May 2016 [6]. However, this suddenly changed in June 2022, when a blog post [7] announced the introduction of *personal achievements*, a new feature whereby *badges* are unlocked in recognition of developers' activities and displayed on their personal profile pages. However, unlike SO, neither the list of available badges is publicly disclosed by GITHUB nor their meaning and the rules to unlock them. Previous research has shown that implementing gamification elements in a collaborative software platform may steer the behavior of developers in unexpected and unwanted directions [8]. We have found evidence that it is not entirely clear to the GITHUB community how to interpret the meaning of personal achievements. In Fig. 1, a user asks the community for help (1) on how to interpret the `Pull Shark` badge (2), and two other users agree with the original poster that the meaning of the badge is ambiguous and not intuitive (3).

GITHUB not only has earned the reputation as the one-stop-shop for software development [9], but has also quickly become a useful platform for job seekers [10–12]. Employers access the GITHUB profile pages of job candidates to supplement resumes and further assess their knowledge, skills, and abilities. Marlow and Dabbish [13] found that the design of GitHub strongly influenced the activity traces available through profile pages, which employers look at during applicant evaluation. In particular, they observed that employers look at the activity traces that provide reliable cues and require less effort to access and evaluate them. Because personal achievement badges are gamification elements that provide quick and low-effort access to activity traces, GitHub users who display them on their profile pages may be unintentionally 'giving off' cues that have unexpected repercussions, for example, to managers assessing them as job candidates and to other OSS developers reviewing their contributions [14,15].

* Corresponding author.
 *E-mail address:* fabio.calefato@uniba.it (F. Calefato).

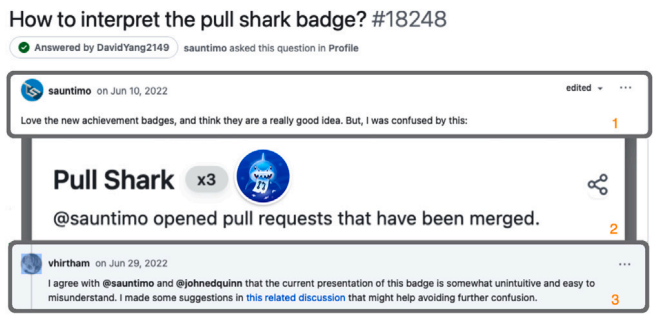

**Fig. 1.** An excerpt taken from a GitHub discussion on the difficulty of interpreting the meaning of badges.

In this paper, we investigate the effects of introducing a new gamification element into GitHub. Specifically, we conduct a mixed-methods study of personal achievement badges (in short, *personal badges* hereafter). We characterize the distribution of badges in GitHub ($RQ_1$), investigate the technical skills, attributes, and personal dispositions of developers they are intended to signal and how well they correlate with these qualities[1] ($RQ_2$), and examine the effects, if any, that the introduction of this feature has had on developers' activities ($RQ_3$). Finally, we gauge how the GitHub community feels about the new feature ($RQ_4$).

After mining longitudinal data from over 6000 GitHub developers, we find that while most of them own at least one badge, the number of those who chose to have a private page or opted out of displaying badges on their profile has been steadily increasing. Furthermore, correlation and regression analyses reveal that, respectively, only one of the five types of personal badges analyzed (i.e., `Starstruck`) provides a reliable signal of the associated attribute (i.e., popularity) and that its introduction has positively affected the number of followers for developers displaying the badge. Finally, the analysis of 54 responses to a survey and 312 community discussions shows that, while some users find badges somewhat appealing and nice to display, most find them unreliable as visual cues of developers' skills and dispositions.

The remainder of this paper is organized as follows. In Section 2, we illustrate the theoretical background of the paper. In Section 3, we present the four research questions. The empirical study is described in Section 4, together with the dataset and the methods used to carry out the investigation. The results are presented in Section 5 and discussed in Section 6 together with related work. Finally, we discuss the limitations in Section 7 and conclude in Section 8.

## 2. Background

### 2.1. Social translucence

One of the reasons behind the success of GitHub is that it has been designed as a 'socially translucent' system [16], which makes socially salient information (e.g., stars, followers) and the activities of participants (e.g., contribution calendar) visible, thus providing the basis for inferences, planning, and coordination. On the contrary, previous OSS forges like SourceForge were generally opaque to social information.

According to Erickson and Kellogg [16], there are three building blocks to designing socially translucent knowledge platforms, namely *visibility*, *awareness*, and *accountability*; visibility and awareness ensure that users modulate their actions appropriately (e.g., seeing who made the latest changes to a file and who is assigned to fix a bug report), and accountability ensures that norms and rules come into play as effective mechanisms of social control (e.g., feeling the pressure of fixing a build-breaking commit).

We argue that profile pages, which show developers' salient personal information, activity traces, and now also personal badges, act as a hub for socially translucent[2] *signals* that support visibility, awareness, and accountability in GitHub.

### 2.2. Signaling theory

When choosing with whom to collaborate, especially online, much of what we want to know about others is not directly observable. Instead, we rely on *signals*, i.e., perceivable features and actions that indicate the presence of qualities of interest [17]. For example, on collaborative development platforms like GitHub we can rely on the total number of code-review comments as a signal of a developer's timeliness and commitment; similarly, a large number of contributed pull requests can signal technical skills and high productivity.

*Signaling theory*, initially developed in the fields of economics [18] and biology [19], is useful to model the relationship between signals and qualities, and to explain how only certain signals can be considered reliable. At its core, signaling theory is fundamentally concerned with reducing information asymmetry between the *signaler* – i.e., the party who possesses some pieces of information/personal attributes and benefits from the actions taken by the other party – and the *receiver* – i.e., an outsider who lacks information and would in turn benefit from making decisions based on it. For example, in biology, a peacock (the signaler) shows off the size and shape of his tail to be selected in favor of alternative partners (the signaler's benefit) by a peahen (the receiver) because a larger tail is a signal of a healthy bird and a better chance of healthy offspring (the receiver's benefit) [19].

However, not all signals are *reliable*. To be considered reliable, the cost of deceiving a signal must outweigh the potential benefits. In terms of reliability, we distinguish two classes of signals. *Assessment signals*, which are inherently reliable because the signal can be produced only if the signaler possesses the indicated quality. For example, being able to lift a heavy weight in a gym is an assessment signal of someone's strength – a person without enough strength would simply not be able. The other class of signals, named *conventional signals*, is not inherently reliable because the link between signal and quality is arbitrary, and is established by social conventions. For example, owning a gym t-shirt is a conventional signal of strength, as it is easily accessible, even if the wearer is weak. These types of signals are widespread in online interaction; for example, the descriptions on online profile pages are conventional signals, which are unreliable because they are more open to deception. Therefore, for signals to be reliable, they have to be *observable* – receivers need to be able to notice them – and they have to be *costly* to produce. In the labor market scenario, where employers lack information on job applicants, high-quality prospective employees differentiate themselves from the other low-quality applicants by highlighting in their resumé their education; holding a degree with honors from a prestigious institution is a signal that is easily observable for recruiters and is also costly to produce and hard to fake without being smart and hard working [18].

---

[1] For the remainder of this document, we employ the term 'qualities' as an abstraction encompassing developers' skills, attributes, and personal dispositions.

[2] Erickson and Kellogg use the adjective *translucent* instead of *transparent* to highlight privacy concerns and the constant tension between showing just enough and too much information on socially augmented platforms. We use the term consistently.

Previous research has shown that easily observable signals conveyed through online profiles in translucent, social coding platforms (e.g., SO reputation points, GITHUB activity graph) can act as proxies of expertise [20,21], awareness [22], and commitment [15,23]. Consequently, here we investigate the extent to which personal badges in GITHUB can be considered reliable as signals (whether assessment or conventional) of desirable yet directly unobservable skills, attributes, and personal dispositions of developers (e.g., timeliness, popularity, desire to collaborate, etc.), useful in the scenarios of open source software development collaboration (i.e., 'people sensemaking') [20] and job seeking [13].

*2.3. Gamification in SE*

Gamification in software engineering is common, especially in the social programmer software ecosystem [14], where several types of elements of game design are often used [24]. Pedreira et al. [25] conducted a systematic mapping study on gamification in software engineering and found that badges, i.e., tokens that users can display only after completing specified activities [26,27], are the second most used gamification element after point-based reputation. For example, SO implements a reward system that combines reputation scores and leaderboards with badges to publicly reward users for their contributions. Papoutsoglou et al. [28] found that gaining badges in SO favors changes in users' attitudes from conservative to more open-minded and eager for new knowledge.

GITHUB has recently implemented personal badges to reward developers for unlocking achievements through their activities on the coding platform. This is the first work to study the effects of badges and the reactions of the community to this new gamification element since its implementation in GITHUB. Previous research on GITHUB, in fact, has focused on badges added by developers to README pages as visual cues that provide quick access to repository and source code statistics [29–31].

Although intended to increase user participation and steer user behavior in desired directions, gamification can have unwanted effects, even resulting in lower engagement, satisfaction, and performance [32]. Moldon et al. [8] examined how the behavior of GITHUB developers changed after the removal of the code streak counters from the activity graphs displayed on their profile pages. They found that long-lasting streaks became less common, thus highlighting the power of gamification as a channel of social influence and raising awareness on the potential consequences of adding gamification elements to collaborative development platforms.

Finally, SO provides a complete list of existing achievements, as well as rules and even community guidelines [33] to detail how to obtain them. However, an undesirable side effect of knowing which achievements are there and how to unlock them is that the behavior of users may change after obtaining a badge. Grant and Betts [34] observed that the number of edits to posts in SO drastically drops once users obtain related achievements. In GITHUB, neither the list of achievements nor the rules for obtaining them are publicly disclosed. We speculate that the choice of not making the list of achievements public may be intended to help avoid the risk of steering developers' behavior toward the game, thus ensuring that the regular coding activity on the platform is rewarded with achievement badges.

*2.4. Summary of key concepts*

The background of this study is grounded in three interconnected concepts: social translucence, signaling theory, and gamification in SE. Social translucence, as exemplified by GITHUB, emphasizes the importance of visibility, awareness, and accountability on collaborative platforms. This concept is closely tied to signaling theory, which explains how observable features (signals) can indicate unobservable qualities, distinguishing between inherently reliable assessment signals and potentially unreliable conventional signals. GITHUB's recent implementation of personal achievement badges represents a confluence of these concepts, potentially serving as signals of developer qualities while also influencing user behavior. However, the impact of such gamification elements is not always straightforward, as evidenced by previous studies on platforms like SO, highlighting the need to carefully consider their implementation and their effects on developer behavior and platform dynamics.

## 3. Research questions

The overall goal of this paper is to characterize the diffusion, effects, and perception of personal badges in GITHUB. Consequently, we define and answer the following four research questions.

*$RQ_1$ – What is the distribution of personal badges on GitHub?*

With the first research question, we want not only to uncover the distribution of unlocked personal badges by type, but also to assess whether and how many developers opted out of displaying them on their public profile page.

*$RQ_2$ – Are personal badges reliable signals of developers' qualities?*

The second research question is based on signaling theory. Because there is no official description of personal badges and how to obtain them, only unofficial resources such as repositories[3] and community discussions,[4] we first seek to understand what developer qualities they are signaling (e.g., timeliness, popularity, desire to collaborate). Then, we assess whether these signals are reliable.

*$RQ_3$ – Has the introduction of personal badges affected developers' activities?*

Although the ultimate benefit of implementing gamification in GITHUB should be the promotion of best practices for collaborative software development, previous research has shown that gamification elements can instead lead to unintended changes in developers' behavior. As such, our objective is to understand the effects of the introduction of personal badges on the activities that developers typically perform in GITHUB. As detailed in Section 5.2, we define activities in terms of several selection and collaboration scenarios that are typical of OSS development.

*$RQ_4$ – How does the GITHUB community feel about personal badges?*

Finally, we investigate the response of the GITHUB community to the introduction and perceived usefulness of personal badges.

## 4. Empirical study

*4.1. Data*

We collected a multidimensional longitudinal dataset of badges and other activities of ~6k GITHUB developers, using the following procedure.

We collected data from a sample of projects, building on our previous study [35] that examined the extent of developer disengagement among GITHUB repository contributors. At the time of writing, GITHUB hosts ~28 million public repositories. Kalliamvakou et al. [36] reported several perils when mining GITHUB, as many repositories do not contain source code or are not software-engineered projects. Following their recommendations, we devised a selection approach that would filter out personal or inactive projects that do not contain source code (e.g., static websites, documentation projects) or do not have a sufficient development activity history (e.g., commits, pull requests). Consequently, we started with the selection of organizations and, First, from the *Topics* section on the GITHUB website, we identified the ten most trending topics. Five to six organizations per topic were selected, resulting in a total of 58 organizations. For each of them, we used the GITHUB API to retrieve the list of members of each organization. The API returned an error when retrieving the members of seven organizations; hence, we ended up with 6022 random developers from 51 organizations. In our sample, we found a balance between volunteers (51%) and paid contributors (49%).[5] Then, for each organization, we selected the 'reference project' (e.g., rails/rails); when unclear (e.g., laravell/framework), we chose the largest in terms of contributors and, if more projects had a similar number of contributors, we chose the one with the most stars. With this approach, we were able to generate a heterogeneous sample of projects that vary in terms of size (contributors, pull requests, LOC), history (age), and programming language. Refer to the supplementary material in the replication package[6] for a breakdown of the selected projects' characteristics.

---

[3] https://github.com/Schweinepriester/github-profile-achievements.
[4] https://github.com/community/community/discussions/28656.

Next, we detail the data collected and measures for each research question. To investigate $RQ_1$ (diffusion of personal badges), because the GitHub API does not support the retrieval of personal badges, we developed a custom scraper in Python. For each user, we scraped the profile page to retrieve the list of unlocked personal badges and the time the achievement was first obtained. GITHUB introduced personal badges in June 2022; we repeated scraping monthly for six months, from June to December. We also counted how many GITHUB developers chose not to have a public profile page or opted out of displaying personal badges from their public profile page.

To investigate the reliability of badges as signals of developers' qualities ($RQ_2$) and the potential effects of badges on developers' activities ($RQ_3$), we collected several repeated measures (see Section 4.2) regarding their development activities and popularity. Regarding development-related activities, we used the GITHUB API to mine the number of Issues and Pull Requests (PRs) opened, closed (merged), and worked on (as assignee) by developers in our sample, the time to close issues and merge pull requests, and the number of commits. As a measure of the popularity of developers in the GITHUB community, we also collected the number of followers. All of these metrics were collected monthly for twelve months, from January to December 2022.

To gauge the perceptions of personal badges in GITHUB ($RQ_4$), we used Google Forms to design an online survey consisting of 15 questions (both closed and open). After collecting basic demographic data, the questions focused on why respondents choose (not) to display badges on their profile, whether they consider them indicators of development skills, and what inferences they make about other developers who display badges on their profile pages. The survey was advertised on social media and posted as a question to the GITHUB community discussions space.[7] Respondents received no monetary compensation. A copy of the survey is available online as supplementary material.[8] Furthermore, to complement the survey and also further assess how the GITHUB community reacted to the introduction of personal badges, we extracted all the discussions from the community/community repository that were listed under the profile category[9] and also contained the keyword badge and/or achievement.

---
[5] We used the approach proposed by Zhang et al. [37]. We analyzed the domain of the email addresses associated with commits. Developers using email addresses from free, general provider domains (e.g., *gmail.com*) were categorized as volunteers. Conversely, developers using email addresses registered to company or organization domains (e.g., *mozilla.com*) were classified as paid contributors.
[6] https://doi.org/10.5281/zenodo.7501582.
[7] https://github.com/community/community/discussions/37346.
[8] https://doi.org/10.6084/m9.figshare.26105695.
[9] https://github.com/community/community/discussions/categories/profile.

## 4.2. Methods

To answer $RQ_1$ (diffusion of personal badges), we compute the number and percentages of badges unlocked by developers in our experimental sample for each month, from June to December 2022, together with the number of developers who opted out of visualizing their badges on the profile page or made their profile page private. In addition, we plot the monthly progression of the number of unlocked badges.

Regarding $RQ_2$ (reliability of badges as signals), we first hypothesize and then search for correlations between the presence of personal badges and differences in the qualities they might signal. We collect several measures, such as the number of followers, commits, and open and closed issues and PRs. After verifying the presence of non-normal distributions with the Shapiro–Wilk test, we use the non-parametric Wilcoxon–Mann–Whitney (WMW) test to compare distributions and Cliff's $\delta$ to assess the effect size.

Personal badges may be correlated with various qualities (i.e., skills, attributes, and dispositions) of developers. However, the analysis performed for the second research question cannot assess whether and how the introduction of personal badges has any effect on developer activities. Consequently, for $RQ_3$ we perform a difference-in-difference (DiD) regression analysis [38], a quasi-experimental approach that uses longitudinal data from observational studies to assess the causal impact of an intervention by comparing changes that occur over time in the outcome variable (e.g., number of commits) between the treatment group (i.e., developers who unlocked the Galaxy Brain badge) and the control group (i.e., developers who did not) [39].

Finally, to answer $RQ_4$ (perceptions of personal badges), we first quantitatively and qualitatively analyze the responses to the survey; then, we perform a thematic analysis of the discussion posts mentioning personal badges. Specifically, we focus on the questions that express positive/negative feedback, report shortcomings, and provide suggestions to improve personal badges.

## 5. Results

### 5.1. $RQ_1$ – Diffusion of personal badges

Table 1 shows the distribution of unlocked badges in our data set as of December 2022. The statistics in the table reveal that the most popular badge is Arctic Code Vault Contributor (4591, 27.02%). This result is 'inflated' because it is among the badges that already existed before the launch of the personal badges in June 2022; in addition, this badge is not earnable anymore, as it was awarded to anyone contributing code to GITHUB in 2020. The other previously existing badges are Mars 2020 Contributor (282, 1.66%), and Public Sponsor (118, 0.69%), unlocked, respectively, by developers who contributed code to the repositories used in the Mars 2020 helicopter mission and by those who have provided sponsorships to projects through the GITHUB SPONSORS program. Because these three badges predate the introduction of personal achievement badges, they were excluded from subsequent analyses.

Regarding the other recently introduced types of badges, the most common (aggregated by tier) is Pull Shark (4762, 28.03%), unlocked by merging a certain number of pull requests. The second most common type of badge in our dataset is Starstruck (2385, 14.03%), which is awarded by owning repositories that receive more and more stars from other users. Pair Extraordinaire (1824, 10.74%) and Galaxy Brain (263, 1.55%) are unlocked, respectively, by coauthoring merged pull requests and by answering questions in project discussions. Finally, YOLO (1324, 7.79%) is the personal badge awarded to those who have merged a pull request without performing a code review. We analyzed a sample of approximately 1900 Pull Requests (PRs) that were merged on the same day their authors obtained the YOLO badge. After discarding 362 PRs (19%) submitted to personal projects, we found that 86% (1316) of the remaining PRs were merged by the authors themselves. This suggests that these contributors are likely maintainers or core developers with write access to the code repository. Because the action that earns developers this badge is a development practice that should not be encouraged, it was also excluded from subsequent analyses. We also discuss the bad coding practice promoted by the YOLO badge during the analysis of the survey responses in Section 5.4.1.

**Table 1**
Categories of personal achievement badges unlocked as of December 2022 and their distribution in our dataset. The gray-hued badges are excluded from empirical analysis.

| Badge | Title | Earnable? | Description | Unlocked (%) |
|---|---|---|---|---|
| | **ACHIEVEMENTS** | | | |
| | Pair Extraordinaire | ✓ | Co-authored a merged pull request | 847 (4.99%) |
| | Quickdraw | ✓ | Closed an issue or a pull request within 5 min of opening | 1441 (8.48%) |
| | Starstruck | ✓ | Created a repository that has 16 stars | 1,122 (6.60%) |
| | Galaxy Brain | ✓ | 2 accepted answers | 192 (1.13%) |
| | Pull Shark | ✓ | 2 pull requests merged | 838 (4.93%) |
| | YOLO | ✓ | Merged a pull request without code review | 1324 (7.79%) |
| | Arctic Code Vault Contributor | ✗ | Contributed code to repositories in the 2020 GITHUB Archive Program | 4,591 (27.02%) |
| | Public Sponsor | ✓ | Sponsoring open source work via GITHUB Sponsors | 118 (0.69%) |
| | Mars 2020 Contributor | ✗ | Contributed code to repositories used in Mars 2020 Helicopter Mission | 282 (1.66%) |

| Badge | Title | Tier | Description | Unlocked (%) |
|---|---|---|---|---|
| | **TIERS** | | | |
| | Pair Extraordinaire x2 | Bronze | Co-authored 10 merged pull requests | 455 (2.68%) |
| | Pair Extraordinaire x3 | Silver | Co-authored 24 merged pull requests | 469 (2.76%) |
| | Pair Extraordinaire x4 | Gold | Co-authored 48 merged pull requests | 53 (0.31%) |
| | Starstruck x2 | Bronze | Created a repository with 128 stars | 598 (3.52%) |
| | Starstruck x3 | Silver | Created a repository with 512 stars | 507 (2.98%) |
| | Starstruck x4 | Gold | Created a repository with 4096 stars | 158 (0.93%) |
| | Galaxy Brain x2 | Bronze | 8 accepted answers | 41 (0.24%) |
| | Galaxy Brain x3 | Silver | 16 accepted answers | 18 (0.11%) |
| | Galaxy Brain x4 | Gold | 32 accepted answers | 12 (0.07%) |
| | Pull Shark x2 | Bronze | 16 pull requests merged | 1,382 (8.13%) |
| | Pull Shark x3 | Silver | 128 pull requests merged | 1,897 (11.17%) |
| | Pull Shark x4 | Gold | 1024 pull requests merged | 645 (3.80%) |

As of December 2022, most of the 6022 developers in our dataset had at least one personal badge displayed on their profile page (4977, 82.65%), only 1 (0.02%) had no personal badge, and 1044 (17.34%) had opted out of displaying badges or having a public profile. Fig. 2 plots over six months the distributions of the number of developers with badges (with preexisting badges excluded) versus those who either opted out of displaying badges on their profile page or without a public profile altogether. We observe an initial increase in the number of users who display badges from June (4825, 80.12%) to July (5163, 85.74%); afterward, we notice a decreasing trend (4452, 73.93% in November) with more and more developers choosing not to display badges, either opting out or because they chose not to have a public profile page (1412, 23.48%).

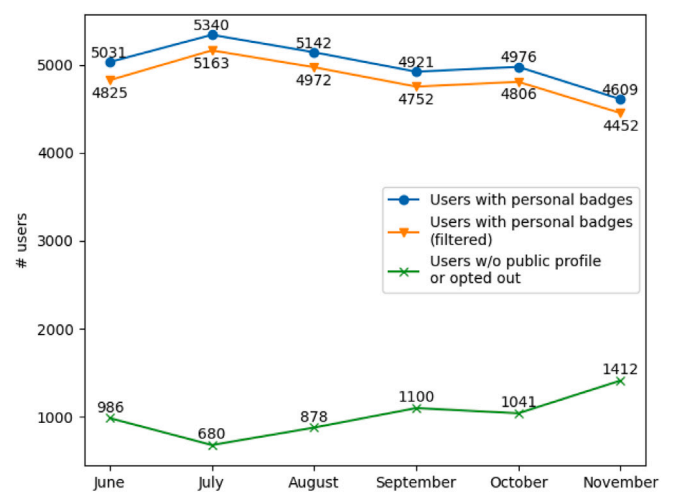

**Fig. 2.** Number of users with badges vs. those who opted out of displaying badges or without a public profile page (the filter refers to the exclusion of preexisting badges.).

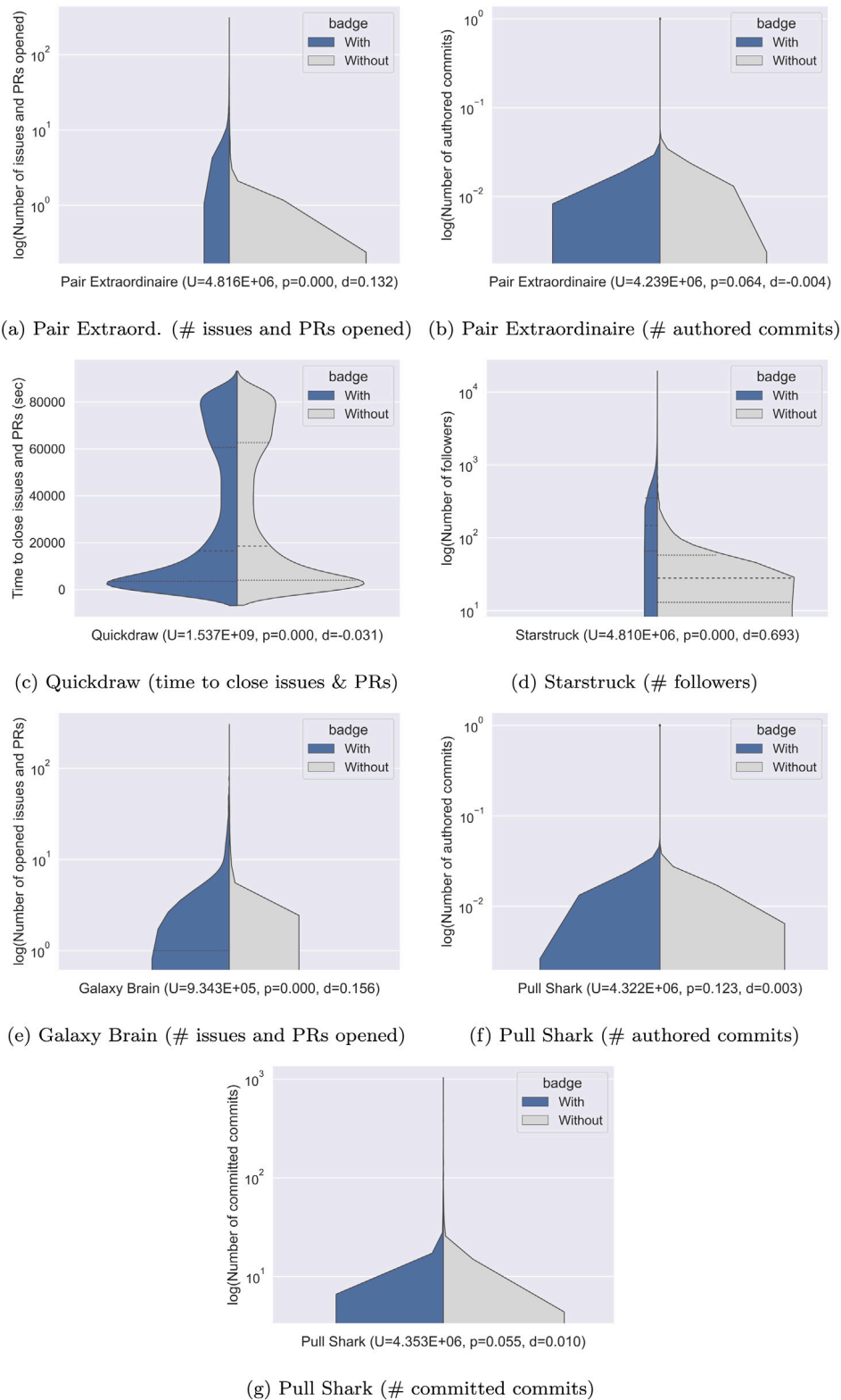

Fig. 3. Distributions of response variables with and without badges. WMW $U$, $p$-value, and Cliff's $\delta$ statistics below each figure.

## 5.2. RQ$_2$ – Reliability of personal badges as signals

**Correlation analysis**. To discover what badges might be signaling, we investigated the associations of their presence with desirable qualities (i.e., skills, attributes, and dispositions) of OSS developers. The results of the correlation analysis between the badges and their hypothesized qualities are reported below. The distributions are illustrated in Fig. 3, where we report the results of the WMW tests with $p$-values. Cliff's $\delta$ values are also reported to gauge the effect size (the magnitude is assessed using the thresholds provided in [40], i.e., $|\delta| <$ .147 '*negligible*', $|\delta| <$ .33 '*small*', $|\delta| <$ .474 '*medium*', otherwise '*large*').

To gauge how GITHUB developers perceive the signals sent by personal badges, we defined the following question in the survey (further described in Section 5.4): "*$Q_{14}$. For each badge, please report the intended signal that you think it sends to others and, if different, the actual signal it conveys to people visualizing it*". As detailed in Section 5.4.1, 54 GITHUB developers participated in the survey. However, for question $Q_{14}$ we received 39 responses, with 8 respondents answering that they had no understanding of the meaning of some badges ($P_{30}$: "*They're supposed to show me some kind of evidence, but I don't know*"). Consistently with the relatively low number of useful answers (31), the analysis of question $Q_{14}$ suggests that the GITHUB community may not have a clear understanding of what signals, if any, personal badges send to others. However, the valid answers show some consistency in describing the perceived signals. After analyzing and grouping the mentions of similar signals in the answers, which we considered only if they appeared at least three times, we hypothesized associations between the perceived signal sent by each personal badge and some code development metrics available through the GITHUB API, as described below.

The `Pair Extraordinaire` badge is awarded to developers who coauthor merged pull requests. According to the clustered survey responses ($n = 6$), owning the badge might signal an increased *desire to collaborate* ($P_{37}$: "*The badge] suggests that they like collaborating with other developers to create some pieces of code*"). Consequently, we tested its correlation with the number of authored commits, as well as the number of PRs and Issues opened, as pull-based development has been popularized by GITHUB to collaborate and integrate developers' contributions to project repositories [41,42]; also, we aggregated the number of PRs and Issues opened, due to their interlinked nature in GITHUB and because opening an Issue (e.g., for reporting bugs or requesting features) is a valuable form of non-coding collaboration [43]. The results of the WMW tests show the existence of a statistically significant difference between developers with and without the badge and the number of Issues and PRs opened (Fig. 3(a), $p < .001$), but with a negligible effect size ($\delta = .132$). Instead, there is no statistically significant difference in the case of the number of authored commits between developers with and without the `Pair Extraordinaire` badge (Fig. 3(b)).

Regarding the `Quickdraw` badge, gained by developers who close Issues and PRs within 5 min, according to the the grouped survey response ($n = 7$) its presence could be a signal of *timeliness* ($P_{14}$: "*It may indicate that you're fast, committed;*" $P_{19}$: "*The word [Quickdraw] says is it... they're quick;*" $P_{40}$: "*I guess it signals you're on time, which means dedication too*"). Consequently, we tested its association with an overall shorter time taken to close Issues and PRs compared to those who do not own the badge. The result (Fig. 3(c)) shows a statistically significant difference in the WMW test, but a negligible effect size ($p < .001$, $\delta < -0.031$).

The most substantial finding is observed for the `Starstruck` badge, which, according to the the similar responses ($n = 9$), is perceived as a signal of *popularity* since it is unlocked when repositories are starred ($P_{41}$: "*It tells others that you are a coding rockstar;*" $P_{50}$: "*[It] means that you're popular, I guess*"). Therefore, we tested the correlation of its presence on developers' profile pages with the number of their followers, a commonly used proxy measure of developer popularity in GITHUB [44]; the results in Fig. 3(d) show that there is a statistically significant difference between the distributions of developers with and without the badge in terms of the number of followers ($p < .001$) and that this difference has a large effect size ($\delta = .693$).

The `Galaxy Brain` badge is unlocked when answers posted to project discussions are accepted. According to respondents ($n = 4$), its presence could signal a greater *willingness to help* projects grow ($P_{8}$: "*Maybe support? That you want to help a GITHUB project and its community beyond 'just' coding?*"). DISCUSSIONS [45] is a new feature for asking questions or discussing topics in GITHUB projects, outside of specific Issues or PRs, in a way similar to Question-Answering sites such as SO. The analogy with SO suggests that developers may be motivated to respond to GITHUB posts for similar reasons, most notably the desire to help fellow programmers [33], contribute to the community [46], and build a reputation [47]. As such, we tested the correlation between its presence and the number of Issues and PRs opened, which again are the main features of GITHUB used by developers to contribute to projects [42]. The results in Fig. 3(e) show that there is a statistically significant difference ($p < 0.001$) and a small effect size ($\delta = 0.156$) between developers with and without the badge in terms of opened Issues and PRs.

Finally, the `Pull Shark` badge is obtained when PRs are merged into a project repository. Based on survey responses ($n = 5$), its presence could represent a signal of *willingness to contribute code* to projects ($P_{29}$: "*As a maintainer, [Pull Shark] gives me the general idea of someone who wants to write code, for their own or others' repos;*" $P_{42}$: "*[It] signals that you're an active code contributor*"). Consequently, we tested its correlation with the number of both authored and committed commits. In the case of authored commits, the results of the WMW tests show a statistically significant difference, but a negligible effect size between developers with and without the badge (Fig. 3(f), $p < 0.01$, $\delta = 0.023$). Regarding committed commits, we observe a lack of statistical significance (see Fig. 3(g)).

**Correlation analysis by tier**. In general, the correlation analysis revealed limited significant results (Table 2, column *All*). Most differences, in fact, are significant but with negligible or small effect sizes, with the exception of the `Starstruck` badge, which exhibits a strong association with the number of followers, as a signal of popularity. One reason for the negligible effect sizes may be that the correlation analysis was performed by aggregating all badge tiers (Table 1), thus distinguishing only between developers *with* and *without* badges. Therefore, it might be possible that the tier-2 (*bronze*), tier-3 (*silver*), and tier-4 (*gold*) badges, which are harder to achieve, might send more reliable signals than the basic tier-1 badges of the same type. Consequently, in Table 2 we report the results of the correlation analysis by tier. However, we observe only a few variations. The difference in the number of followers between developers without the `Starstruck` badge and those who unlocked the basic tier is significant, but with a negligible effect size ($\delta = -0.031$); on the contrary, the effect sizes remain large for the other tiers. Furthermore, for the `Galaxy Brain` badge, we observe a significant difference with a medium effect size ($\delta = 0.341$) only for the *silver* tier.

### 5.3. $RQ_3$ – Effects of personal badges on developers' activities

This section presents the results of the difference-in-difference (DiD) regression analysis. The fundamental assumption for applying a DiD regression is that in the absence of treatment (i.e., before the introduction of personal badges) the dependent variable trend would be the same in both the treatment group (i.e., the developers who will eventually unlock a given type of badge) and the control group (i.e., the developers without the badges). Developers who opted out of displaying badges on their profile page and chose to keep the profile private were excluded from the following analysis. In Fig. 4(a), using boxplots we visually confirm the *parallel* (or *common*) *trends* assumption for the `Starstruck` badge; Fig. 4(b) instead shows a violation of the premise for the `Quickdraw` badge, which is therefore excluded from this analysis. The other badge excluded from this analysis for the same reason is `Pull Shark`, in terms of both the number of authored and committed commits. Refer to the supplementary material in the replication package for the other figures.

A difference-in-difference regression model is built to estimate equations such as (1)

$$y_{it} = \alpha + \beta B_i + \gamma BA_t + \delta(B \times BA)_{it} + \epsilon_{it} \quad (1)$$

where $y_{it}$ is the dependent variable (outcome) for developer $i$ at time $t$, $B_i$ is the binary variable for developers who own a badge (the treatment group, i.e., $HasBadge_i = 1$) or not (the control group, i.e., $HasBadge_i = 0$), and $BA_t$ is a time dummy that switches on for observations obtained after June 2022 when personal badges became available (i.e., $BadgeAvail_t = 1$ when $t > June$, otherwise $BadgeAvail_t = 0$). Finally, $\epsilon_{it}$ is the residual term.

Table 2

Results of the WMW test for each badge, aggregated and by tier. Significant results with medium or large effect sizes are shown in bold.

| Badge | All | | Basic tier (x1) | | Bronze tier (x2) | | Silver tier (x3) | | Gold tier (x4) | |
|---|---|---|---|---|---|---|---|---|---|---|
| | WMW U | $\delta$ | WMW U | $\delta$ | WMW U | $\delta$ | WMW U | $\delta$ | WMW U | $\delta$ |
| Pair extraordinaire (# Issues & PRs opened) | $4.82 \times 10^{6}$*** | 0.132 | $2.11 \times 10^{6}$*** | −0.031 | $1.22 \times 10^{6}$*** | 0.146 | $1.34 \times 10^{6}$*** | 0.218 | $1.48 \times 10^{5}$*** | 0.229 |
| Pair extraordinaire (# authored commits) | $4.57 \times 10^{6}$*** | 0.073 | $2.07 \times 10^{6}$*** | −0.031 | $1.16 \times 10^{6}$*** | 0.085 | $1.21 \times 10^{6}$ | 0.101 | $1.34 \times 10^{5}$* | 0.115 |
| Quickdraw (time to close Issues & PRs) | $1.54 \times 10^{9}$*** | −0.031 | $1.54 \times 10^{9}$*** | −0.031 | – | – | – | – | – | – |
| Starstruck (# followers) | $4.81 \times 10^{6}$*** | **0.693** | $2 \times 10^{6}$*** | −0.031 | $1.26 \times 10^{6}$*** | **0.755** | $1.17 \times 10^{6}$*** | **0.880** | $3.77 \times 10^{5}$*** | **0.958** |
| Galaxy brain (# Issues & PRs opened) | $9.34 \times 10^{5}$*** | 0.156 | $6.71 \times 10^{5}$*** | −0.031 | $1.51 \times 10^{5}$*** | 0.236 | 75,607.5*** | **0.341** | 36,937.5 | 0.072 |
| Pull shark (# authored commits) | $4.41 \times 10^{6}$** | 0.023 | $7.48 \times 10^{5}$ | −0.031 | $1.25 \times 10^{6}$ | 0.002 | $1.78 \times 10^{6}$*** | 0.032 | $6.3 \times 10^{5}$*** | 0.082 |
| Pull shark (# committed commits) | $4.33 \times 10^{6}$ | 0.004 | $7.44 \times 10^{5}$** | −0.031 | $1.24 \times 10^{6}$ | −0.008 | $1.74 \times 10^{6}$ | 0.007 | $6.08 \times 10^{5}$*** | 0.045 |

\* $p < 0.05$.
\*\* $p < 0.01$.
\*\*\* $p < 0.001$.

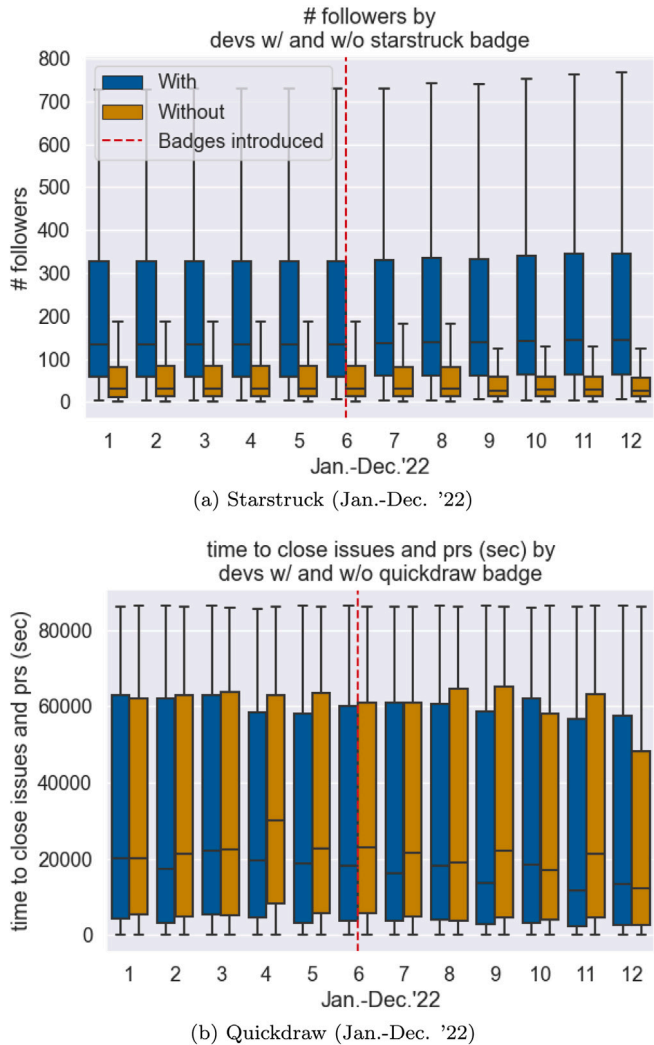

**Fig. 4.** Examples of visual confirmation (a) and violation (b) of the assumption of parallel trends before the introduction of personal badges.

The resulting DiD model is an interaction model interpreted as follows [38]. The intercept estimate $\alpha$ is interpreted as the mean of the outcome (i.e., the dependent variable $y$) for the control group (i.e., the developers without the badge) in the month(s) before the introduction of the personal badges feature in GITHUB (i.e., $HasBadge_i = 0$ and $BadgesAvail_t = 0$). The coefficient $\beta$ of $HasBadge$ is the expected mean change in the outcome $y$ between the treatment and control groups in the pre-treatment period ($HasBadge_i = 1$ and $BadgesAvail_t = 0$); this can be viewed as the 'baseline difference' in the outcome variable between the two groups before treatment. The coefficient $\gamma$ of $BadgesAvail$ is the expected mean difference in $y$ before and after the introduction of the personal badges among the control group; the main effect for $BadgesAvail$ is the effect of the simple passage of time in the absence of the treatment. Finally, the estimated coefficient $\delta$ of the interaction term is an estimate of the treatment effect. This is the DiD coefficient and the focus of this analysis because the interaction is to test whether the expected mean change in the outcome $y$ before and after the introduction of the new feature is different for the treatment and control groups, respectively, the developers with ($HasBadge_i = 1$) and without ($HasBadge_i = 0$) the badge of interest.

Table 3 presents the results of the DiD regressions using the latest data snapshot collected in December 2022. They show that the interaction term is significant only for the Starstruck badge, on which we focus on the presentation of findings. The DiD coefficient $\delta$ is significant and different from zero (0.169); this means that the introduction of the new feature of the Starstruck badge, which counts the number of starred repositories of a developer, caused an increase in the average number of followers; specifically, because the regression is in logs, the average count of followers in the post-treatment time window is 18.4% (i.e., $(e^{coef} − 1) \times 100$) higher than it would have been without the introduction of the Starstruck badge in GITHUB [48]. In addition, we observe that the coefficient $\beta$ (treatment group) is significant and different from zero, which means that in the pre-treatment time window the developers in the treatment and control groups had a different number of followers; specifically, the developers with the badge had on average 299.9% more followers than those without it. Finally, the coefficient $\gamma$ is also significant and different from zero (−0.1), meaning that the average number of followers in the control group from the pre-treatment to the post-treatment time window decreased by 9.5%.

### 5.4. RQ4 – Community perception of personal badges

#### 5.4.1. Survey analysis

We received 57 responses. The survey analysis was carried out by the first two authors. They analyzed the survey results reading and annotating the responses in a spreadsheet to identify interesting excerpts from open questions and highlight commonalities. After that, the entire team discussed and consolidated the extracted excerpts.

Regarding demographics ($Q_1$–$Q_4$), participants in the survey reported having on average 9 years of experience (min. 1, max. 31, median 8). The participants also maintain an average of 6 non-personal, open source software repositories (min. 0, max. 18, median 4) and have contributed to ~22 public repositories (min. 1, max. ~100, median 15). Finally, they reported having an average of ~28 followers (min. 0, max. ~100, median 22). Three respondents were disqualified because they reported ($Q_5$) that they had opted out of displaying badges (1) or did not have a public profile page (2). They motivated their choices ($Q_{15}$) by saying that they "*simply don't like the idea of badges*" ($P_{22}$) and that GITHUB "*may be a social coding site but it's not (or shouldn't be or become) like social media*" ($P_{33}$). The other users who reported keeping the badge

Table 3
Results of the difference-in-difference regression for each badge. Significant results are highlighted in bold.

Pair extraordinaire, log(y=no. issues and PRs opened)

|  | Coef (S.E.) | [0.25 | 0.75] |
|---|---|---|---|
| Intercept ($\alpha$) | **0.295 (0.142)*** | **0.015** | **0.574** |
| HasBadge ($\beta$) | 0.054 (0.198) | −0.335 | 0.443 |
| BadgeAvail ($\gamma$) | **0.913 (0.169)*** | **0.581** | **1.245** |
| HasBadge:BadgeAvail ($\delta$) | 0.374 (0.229) | −0.077 | 0.825 |

Adj. $R^2$ = 0.212
Treatment $n$ = 250, Control $n$ = 170

Pair extraordinaire, log(y=no. authored commits)

|  | Coef (S.E.) | [0.25 | 0.75] |
|---|---|---|---|
| Intercept ($\alpha$) | **8.972 (0.053)*** | **8.904** | **9.041** |
| HasBadge ($\beta$) | −0.107 (0.067) | −0.239 | 0.025 |
| BadgeAvail ($\gamma$) | **−0.200 (0.040)*** | **−0.277** | **−0.122** |
| HasBadge:BadgeAvail ($\delta$) | 0.072 (0.076) | 0.344 | −0.077 |

Adj. $R^2$ = 0.004
Treatment $n$ = 250, Control $n$ = 170

Starstruck, log(y=no. followers)

|  | Coef (S.E.) | [0.25 | 0.75] |
|---|---|---|---|
| Intercept ($\alpha$) | **3.618 (0.011)*** | **3.596** | **3.639** |
| HasBadge ($\beta$) | **1.386 (0.016)*** | **1.354** | **1.418** |
| BadgeAvail ($\gamma$) | **−0.100 (0.016)*** | **−0.131** | **−0.070** |
| HasBadge:BadgeAvail ($\delta$) | **0.169 (0.023)*** | **0.124** | **0.213** |

Adj. $R^2$ = 0.228
Treatment $n$ = 30,374, Control $n$ = 26,2671

Galaxy brain, log(y=no. issues and PRs opened)

|  | Coef (S.E.) | [0.25 | 0.75] |
|---|---|---|---|
| Intercept ($\alpha$) | **0.322 (0.104)**** | **0.117** | **0.527** |
| HasBadge ($\beta$) | 0.007 (0.399) | −0.778 | 0.792 |
| BadgeAvail ($\gamma$) | **1.166 (0.121)*** | **0.929** | **1.404** |
| HasBadge:BadgeAvail ($\delta$) | −0.135 (0.436) | −0.991 | 0.721 |

Adj. $R^2$ = 0.187
Treatment $n$ = 46, Control $n$ = 375

* $p < 0.5$
** $p < 0.01$.
*** $p < 0.001$.

feature enabled ($Q_6$, $n$ = 54), cited various motivations for doing so. The primary reason was curiosity (19 respondents), followed by having observed badges on other developers' profiles (18 respondents), the aesthetic appeal (12), as a suggestion by fellow GitHub users (3), and in recognition of the feature's utility in conveying valuable information (2). In the following, we analyze the remaining valid responses.

Question $Q_7$ ($n$ = 53, see Fig. 5(a)) aimed to assess whether participants consider the presence of personal badges an indicator of their own coding skills in general. Most of the respondents either disagreed ($n$ = 8) or disagreed strongly ($n$ = 24). Then, questions $Q_8$–$Q_{10}$ ($n$ = 53) asked participants to report on the effects that badges can have on collaboration with other developers. The results are also reported in Fig. 5. Although the distribution of responses is varied, overall the figure shows that about half of the respondents feel that badges cannot be an indicator of others' ability and that they are not influenced by badges' presence during collaboration with others. To deepen our analysis, we investigated whether more experienced developers have a different perception of personal badges as indicators of coding skills and their effect on collaboration. Therefore, we filtered the responses to questions Q7–Q10 to include only participants reporting above median scores for the demographic questions (Q1–Q4) addressing development experience, the number of non-personal open-source software (OSS) projects maintained and contributed to, and the number of followers. The proportionally higher number of disagreements in Fig. 5(b) shows that experienced developers are more skeptical about the potential usefulness of personal badges in GitHub.

Question $Q_{11}$ asked the respondents to justify their previous responses. They were critical of the current implementation of badges. Participant $P_7$ found that "*these badges can be achieved too easily, not like in Stack Overflow*", and $P_2$ that there are too few of them ("*we all have almost the same badges*"); others added that they fail to adequately capture developers' experience – e.g., "*I have 30+ years of development experience and almost none is visible through [badges]*" ($P_1$), "*I have lots of experience and almost no badges*" ($P_8$). Furthermore, $P_5$ responded by asking a provocative question: "*Are developers with these badges better developers? Vice versa and more critical: Is a developer bad just because they don't have some of these badges?*"

Next, we examine how respondents feel about specific badges. Question $Q_{12}$ ($n$ = 41) asked participants to indicate the most relevant badges. The answers are clustered around two main groups. The first group contains answers (15) from users who feel that none of these badges can tell us anything useful about others. The other group of remaining answers (26) instead believes that the badges give information about who unlocks them. Specifically, 6 respondents mentioned the badge GALAXY BRAIN (awarded to those who have accepted answers in project discussions) because it reflects "*how serious [one is] in the community*" ($P_{28}$). A group of 11 participants mentioned the STARSTRUCK badge (unlocked by receiving project stars) because it gives tangible "*evidence of recognition and interest by the community*" ($P_{17}$). The remaining respondents (9) highlighted the importance of PAIR EXTRAORDINAIRE (awarded to those who coauthor merged pull requests) because the badge may reflect "*a positive attitude [toward] collaboration with others*" ($P_{19}$).

Finally, question $Q_{13}$ ($n$ = 42) asked participants to indicate any missing badge, in addition to those available. The suggestions converged on the following ideas for new badges that would display: (i) the years of work in a project, (ii) how many repositories you are a maintainer of, (iii) how many repositories you have contributed to, (iv) more badges in general, and (v) more badges that are challenging to unlock.

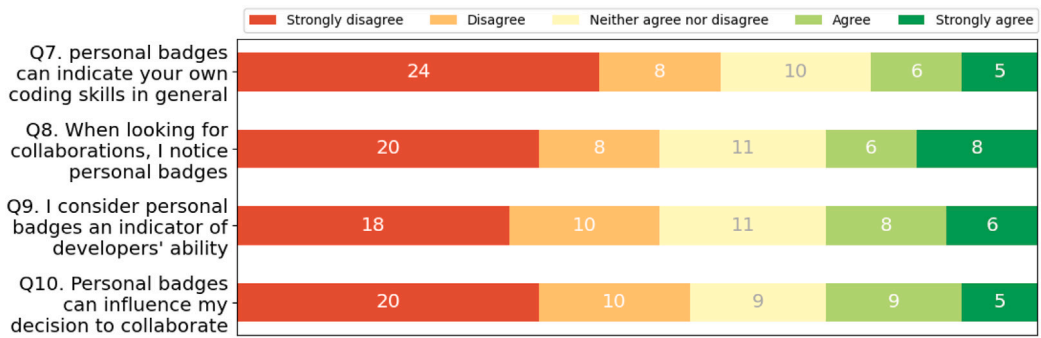

(a) All survey respondents.

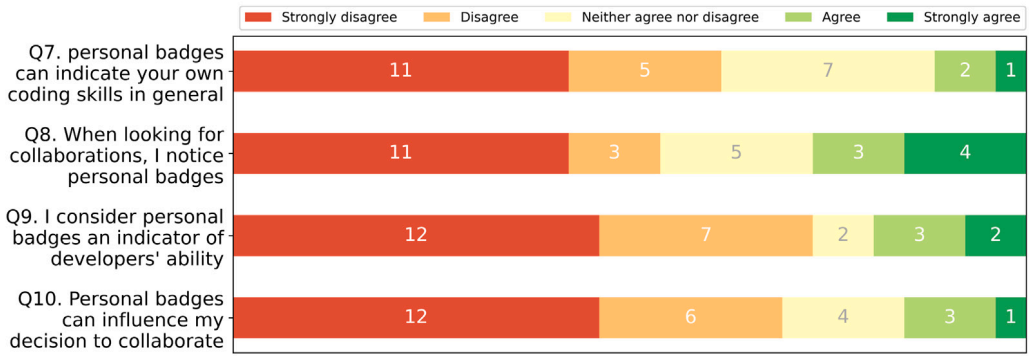

(b) More experienced respondents only.

**Fig. 5.** The extent to which participants find personal badges to be an indicator of coding skills ($Q_7$) and their perceived effects on collaboration ($Q_8$–$Q_{10}$).

*5.4.2. Community discussions analysis*

Overall, we extracted all the Q&A threads related to personal badges from June (i.e., the launch of the new feature) to December 2022 and retrieved 312 questions.

Subsequently, we performed a thematic analysis. We divided the entire set of 312 posts into three subsets of approximately 100 elements; then, the first two authors independently analyzed the first subset. We identified common and different categories. The differences were resolved through discussions with the entire team. For the two other subsets, we followed the same approach, with changes to the coding schema propagated back to the previous batch(es) if necessary.

Thematic analysis revealed four main themes (see Table 4): *Info requests*, *Feedback*, *Improvements*, and *Other*. Next, we describe each theme and the corresponding codes, exemplifying the most relevant concepts with excerpts.

The prevalent theme that emerged from our analysis is the request for information about personal badges — namely, *Info requests*. This theme encompasses 122 distinct questions, almost twice as many as the other two categories. Arguably, the prevalence of questions on this theme is explained by the choice of GITHUB not to disclose explicit information on personal badges; this choice has led several users to seek unofficial information on the community forum. Within *Info requests*, the most frequent code is *How to get a badge*, assigned to 79 questions regarding the requirements needed to earn a specific badge (e.g., *"How do you get the Galaxy Brain Badge?"*). Similarly, 17 more questions were coded as *How to get badges*; these represent requests for help on how to earn more badges in general (*"How to get a new badge on GitHub?"*). The remaining 26 questions from this theme were coded as *General info*; these are generic requests for information about the badges feature, the most common being the full list of the currently available personal badges in GITHUB (e.g., *"How can I see all available badges in GitHub? Please Help!!"*).

The next identified major theme is the *Improvements* category, which includes 60 questions and two codes. Under *Badge(s) proposals*, we collected 22 questions that represent requests for the addition of new personal badges. For example, three users recommended adding an additional badge to reward code-review activity (e.g., *"I've learned a lot from code reviewing others' code and being code reviewed by others"*); furthermore, three other users proposed the addition of organization-specific badges (e.g., *"to celebrate the first PR merge of a new employee"*). Some users even suggested creative names for their new badge ideas, such as the *"Issue Muncher"* badge, to keep track of *"how many issues one has helped close through linked PRs"*, and the *"Bug Hunter"* badge, to reward bug reporting. Also, a user suggested that the personal badge feature be extended to showcase actual professional achievements, such as the Linux Foundation Certified Kubernetes Administrator and Google Certification Academy certificates. Finally, a couple of additional questions with this code asked how to suggest new types of badges, showing that some users are willing to participate in the definition of future personal achievements. Regarding the *UX suggestion* code, we identified 40 threads. In particular, 21 questions concerned a feature request, the most popular (8 questions) being the possibility to selectively hide/show the earned badges on user profile pages; other recurring feature requests are: enabling custom badge ordering in profiles (2), taking historical data into account when defining badge unlocking rules (2), and enabling a summary view displaying all possible achievements (2). Several UX-related questions (11) were posted to criticize the graphical appearance of badges (e.g., *"These new badges are too cartoonish"*, *"The designs could be more professional"*) or suggest how to improve it (e.g., *"Make the old badges look like the new ones"*).

The most significant theme in relation to RQ4 is *Feedback*, which includes the codes *Positive*, *Negative (in general)*, and *Negative (specific)*, for a total of 53 questions. Although some questions (12) report positive feedback, in the form of short and generic statements of appreciation for the feature of personal badges, most of the questions collected within this category were coded as negative feedback, either generic (25 questions) or feature-specific (16 questions). Regarding generic negative feedback, several users reported disregarding the new feature, and 17 out of 25 asked how to opt out. Frequent causes for the negative feedback were: i) aversion to the gamification of a professional environment like GITHUB and the *"childish"* style of badges (11 questions) – e.g., *"[they] look like Candy Crash Saga. These weird cartoon emoji are massive, and distract the eye from actual content"*; ii) skepticism about the ability of personal badges to convey developers' skills (2 questions) – e.g., *"unnecessary gamification that doesn't help assess the performance/dedication/talent/achievement of a developer"*; iii) concerns about the potential negative effects of badges on user behaviors (5 questions) – e.g., *"PRs created not for the purpose of making a meaningful contribution, but simply to get another badge checked off their list"*., *"[achievements] set up terrible incentives for folks to create PRs, not for the purpose of making a meaningful contribution, but simply to get another badge checked off their list"* – and the whole community – e.g., *"[Badges] risk becoming a driver for ivory tower superiority and community tribalism, which is something the developer community already struggles with"*, *"I'm not listening to your feature request because I have badges and you don't"*.

**Table 4**
Results from the thematic analysis of questions.

| Category | Code | Frequency |
|---|---|---|
| Info requests (122) | General info | 26 |
|  | How to get badges | 17 |
|  | How to get a badge | 79 |
| Feedback (53) | Positive | 12 |
|  | Negative (in general) | 25 |
|  | Negative (badge-specific) | 16 |
| Improvements (60) | Badge(s) proposal | 23 |
|  | UX suggestion | 37 |
| Other (77) | Technical issue | 42 |
|  | Miscellaneous | 4 |
|  | Excluded | 31 |

Regarding badge-specific negative feedback, most complaints related to the badge YOLO (7 questions), which is perceived by several developers as a source of shame and an unfair mark to obtain, especially if earned within single-person projects (e.g., *"I feel having the YOLO badge does not send the right message, especially to recruiters"*). Another common criticism (4 questions) concerned the unclear requirements for unlocking the Pull Shark badge, sometimes perceived as biased (e.g., *"Pull Shark achievement should work [only] for repositories I don't own. [...] we will always merge our own pull requests, but not everyone will merge our pull requests"*).

Finally, the theme *Other* gathered 77 questions, mostly coded as requests of support for a *Technical issue* (e.g., *"Pull Shark shows the wrong merged PR number for me"*), and 4 questions coded as *Miscellaneous*. In addition, we excluded 31 questions when their content was empty, deemed off topic, or not in English.

## 6. Discussion

In this paper, we studied personal achievement badges, a new gamification feature introduced in GITHUB, with unknown effects.

**Research questions**. We answered four research questions. First ($RQ_1$), we explored the diffusion of badges among a sample of 6022 GITHUB developers as of December 2022 (Table 1) and their evolution over six months, since their introduction in June 2022 (Fig. 2). We found that all developers except one own at least one badge and that the number of those who opted out of displaying badges and chose to make their profile pages private has been steadily increasing. Then ($RQ_2$), we investigated whether the ownership of badges is associated with signaling certain hypothesized developers' qualities (i.e., skills, attributes and dispositions). The initial results of our exploratory analysis (Fig. 3) showed only a large difference for developers owning the badge Starstruck, gained by those who get their repositories starred and a signal of increased popularity. Furthermore, regression analysis ($RQ_3$) showed that the only effect of implementing personal badges in GITHUB has been a large increase in the number of followers of developers who own the Starstruck badge (Table 3). Finally ($RQ_4$), the analysis of the survey responses showed several shortcomings in the current implementation of badges, such as the limited types of badges currently existing and the lack of those more accurately reflecting contributions to projects and years of experience. Besides, the analysis of community discussions about personal badges revealed that GITHUB users are willing to know more about badges and the requirements set out for earning them; also, part of the community is willing to participate in the definition of future achievements. However, the discussions analyzed showed a clear prevalence of negative opinions on badges, highlighting the skepticism of the community about their ability to adequately convey developers' skills and a general concern about the potential drawbacks of GITHUB gamification.

**Badges as signals**. The results of the user survey confirmed that GITHUB developers do look at profile pages and are aware of the elements therein—so much so that several even decided to opt out of displaying badges. As such, these pages have the potential to act as hubs that make socially translucent signals visible and readily available, and support awareness of collaborators' behavior [16]. This result is consistent with those reported by Shami et al. [20] who used the signaling theory [17] as a conceptual framework to investigate how users of online social platforms rely on digital artifacts for people sensemaking, i.e., use portions of profile pages as proxies used to infer unknown coworkers' expertise.

However, the visibility of personal badges is not sufficient to ensure their reliability. According to signaling theory, to be reliable, assessment signals must be both observable and costly to produce. Our findings suggest that most of the currently available personal badges, although they reflect the properties possessed, do not send reliable assessment signals because they may not be costly enough to produce. The only exception to general unreliability is the Starstruck badge signaling popularity. One potential explanation for this finding is that the badge is awarded to a developer as a result of the interest of others (stars) in their repositories: creating a repository implies an effort broader in scope, if not harder, as compared to other badges, which require the developers to take one action (e.g., answer a question, merge a pull request). This result is somewhat consistent with those reported by Trockman et al. [29], who found that popularity-related repository badges are more reliable as assessment signals. One major difference with their work is that repository badges are *chosen*—maintainers select what they intend to signal by adding them to README pages (e.g., the count of downloads to signal popularity, up-to-date dependencies to signal security). Conversely, personal badges are a truer gamification element – they are unlocked and displayed on personal pages – and the only control developers have over them is to not show them altogether.

**Badges as a gamification mechanism**. According to Hunter and Werbach [2], users of a gamified environment go through a journey, from *onboarding* to *scaffolding* and, finally, *mastery*. Implementing game mechanisms such as badges to gamify a software development environment is an iterative and far from trivial undertaking [49]. There have been so far several attempts to propose frameworks for the gamification of Software Engineering activities [50,51] and methods to build gamified software [52]. Our analyses suggest that the leveling process of personal badges in GITHUB needs to be adjusted, especially during the onboarding and mastery stages. On the one hand, some junior developers among the survey respondents reported that it *"felt nice"* to discover the first badge on their profile. However, the easier-to-achieve onboarding badges (i.e., YOLO and Quickdraw) are obtained through practices that are not to be encouraged in software development (i.e., closing issues and PRs quickly or without code review) – in the community discussions, some users referred to these badges as "*shameful*" and "*trivial*". In contrast, more senior developers argued that they feel that current badges are "*too easily achieved, not like in Stack Overflow*" and complained about the lack of badges reflecting their years of experience and project contributions ("*They don't give the impression that I'm a professional developer, but that I'm a novice coder who's just finished my first project*".).

**Implications for GitHub platform designers**. The negative feedback collected from survey responses and discussion analysis highlight some criticalities in the current design of personal badges. GitHub platform designers should consider using the following insights to make informed decisions about future improvements.

In our study, we found an increasing number of developers (up to ~23%) who made their profile page private or opted out of displaying personal badges. Although neither the survey nor the discussion analyzes helped us understand the motivations, a possible explanation is that GitHub developers are aware that the elements on their profile page may influence the processes of impression formation [14,21] and people sensemaking [20]. For example, more and more employers look at the information gathered from social networking sites during the hiring process to assess job applicants [53], and their image on social media has an influence on the chance of being hired [54]. However, previous research found that, although conscious, only few GitHub developers engaged in efforts to clean up their profile page because they were deemed too costly [13]. Our findings show that GitHub developers may have become more aware over time and that the decision to allow them to easily opt out of displaying personal badges is a step in the right direction. However, GitHub platform designers should also consider implementing the ability to selectively hide/show badges on user profile pages – a feature requested by several survey participants; this might alleviate the discontent caused by badges like `YOLO` and `Quickdraw`, which mark and publicly expose undesirable and unprofessional user behaviors.

In terms of usability, platform designers should also consider making the progress toward badge achievement more explicit, especially in the case of debated badge requirements (e.g., those of the `Pull Shark` badge), and graphic designers should also revise the appearance of badges to give them a more professional and consistent 'look and feel.'

Furthermore, the current implementation of badges in GitHub appears underdeveloped. First, as discussed earlier, the current number of badges and the implementation of the leveling system do not seem to fit well with the onboarding and mastery stages typical of a gamified environment. As such, we suggest platform designers consider adding more: i) *onboarding badges*, that are appealing to newcomers (e.g., junior developers) without fostering bad software development practices such as no or quick code reviews; ii) *mastery badges*, that reflect long-running, broad-in-scope achievements that are appealing to and feel 'earned' by more experienced, senior developers – who have been found to be intrinsically less motivated by gamification incentives [55]. Second, platform designers should consider rolling out more of them all at once rather than slowly, as some users complained that they "*all have almost the same badges*". Sarma et al. [56] developed Visual Resume, a tool that aggregates activity traces collected from GitHub and SO to aid employers in the hiring process; they found that managers and other technical personnel prefer cues in aggregated form (e.g., activity summaries) and community-generated endorsements (e.g., reputation) to inform hiring decisions and form impressions. Their findings may provide platform designers with useful ideas to implement more reliable personal badges, such as having quick access to features indicating whether commits are central to the code base and whether they make significant changes rather than tweak/fix code. Third, platform designers should further reflect on the unintended side effects of the decision not to provide any official documentation regarding badges. Our findings show that this decision implies that no clear indication is given of what cues badges are intended to signal beyond their name and image, with a possible impact on impression formation and people sensemaking during collaborative development [14,21]. At the time of writing, two more badges have been discovered, namely `Heart On Your Sleeve` and `Open Sourcerer`, but the community still has not confirmed what they mean or how to achieve them.[10,11]

Finally, we urge platform designers to reflect on the potential side effects that displaying personal badges may have on the hiring process, as revealed by our analysis of community discussions ("*A potential employer glancing at my profile, perhaps unfamiliar with this [...] feature, might assume a negative impression of me*"). Although GitHub is designed as a platform to support collaborative software development, employers and job seekers also use it as a recruitment tool [57,58]. Marlow and Dabbish [13] analyzed the signals given off by the activity traces displayed on GitHub profile pages through the lens of signaling theory. After interviewing managers and developers who use GitHub as part of the hiring and job application process, they found that interviewees valued the insights provided by GitHub accounts as more reliable and verifiable than a static list of individuals' skills. In fact, according to Walther et al. [59], when forming an impression, the information provided by third parties (i.e., GitHub in this case), is more reliable than self-reports. However, Marlow and Dabbish found that the reliability of the activity traces as signals varied and that their use was directly related to the evaluation cost. In particular, they found that activity traces send reliable assessment signals when they reflect the (i) *history of activity* (e.g., commitment to an open source project over years) and (ii) *networking* (e.g., contributing to high-status OSS projects). However, employers only chose to look at those activity traces that are easy to verify quickly. Hence, consistently with this evidence and signaling theory, we suggest that platform designers overhaul personal badges so that they provide quick access to aggregated traces of activity history and collaboration in high-status projects, which are costly to produce for job seekers and, at the same time, affordable for employers to verify.

## 7. Limitations

First, in this study, we analyzed the activities of a sample of more than 6000 GitHub developers retrieved from 51 organizations. Because GitHub currently hosts a steadily increasing number of OSS projects that are the result of the contributions of millions of developers,[12] we acknowledge the potential threat to external validity. However, we emphasize that our study is the first to analyze personal badges; it was initiated in a timely way as soon as the new feature was released in June 2022 and completed within six months after its release. As such, we argue that the novelty and timeliness of this work compensate for the limited number of sampled developers. Furthermore, we point out that this is a study that compares the activity and profiles of developers *before* and *after* the introduction of the features of personal badges and therefore it was possible to conduct it only because we already had the necessary data collected for the *before* stage as part of a previous study. In other words, without the already available data on the ~6k developers – limited or not – this study would simply not have been possible. We also point out that the sampled developers contribute to large and active software-engineered projects that were carefully mined from GitHub following the recommendations provided in [36]. Finally, given the exploratory nature of our study, we argue that the current findings help us to provide an initial understanding of both the effects of badges and the developers' perception of the new gamification element introduced in GitHub.

---

[10] https://github.com/drknzz/GitHub-Achievements.
[11] https://github.com/github-profile-achievements/english.
[12] https://octoverse.github.com#the-world-of-open-source.

We also acknowledge the somewhat limited number of survey participants (54). However, to counteract this limitation and properly answer RQ4 (gauge community perception of personal badges), we complemented the survey analysis with the thematic analysis of more than 300 discussions where GITHUB community members shared their opinions about the new feature.

Another limitation is that in our quantitative analyses (RQ1-3) we only analyzed the activities of the sampled developers in public repositories. Developers can choose to also include their activities in private repositories as contributing to unlocking personal achievements. However, this is information that is impossible to retrieve and therefore represents an intrinsic limitation that our study shares with all works that rely on data retrieved from public repositories GITHUB. For RQ3, although we verify the parallel trends assumption, there are additional threats to validity that warrant consideration when applying the difference-in-difference regression analysis to infer causal relationship. First, selection bias may exist if there are systematic differences between developers who earn badges and those who do not. However, this risk is mitigated by our diverse sample of developers from various organizations. Second, omitted variable bias could occur due to unaccounted time-varying factors affecting developer behavior. However, our comprehensive data collection on multiple developer activities helps minimize this risk. Lastly, the timing of the effects may not be fully captured in our analysis period. However, our six-month observation window after the introduction of the badge likely captures immediate and short-term effects, which are valuable for understanding the initial impact of this gamification feature. Future studies could extend this time frame to capture potential long-term effects.

Finally, because GITHUB provides no official information on personal badges and how to unlock them, all the descriptions provided in this study regarding badges have been inferred by community members and contributed to public repositories. One potential limitation affecting correlation and regression analyses is that we hypothesized the cues that personal badges might send as signals based on the perceptions reported by survey participants in their responses. Although we considered only the perceived signals that were reported by two or more respondents, alternative hypotheses not included in the DiD regression analysis are still possible. Still, we notice that the potential 'arbitrariness' of the cues sent by badges is an intrinsic limitation of the current implementation of the available personal badges, which lack official documentation.

## 8. Conclusions

In this paper, we analyzed the introduction of a new gamification feature in GITHUB, namely personal achievement badges. We studied the distribution of badges, the signals they send, and the effects of their introduction. We also collected evidence of the reaction of the developers' community to the new feature. We concluded by providing recommendations to the designers of the GITHUB platform on how to improve personal badges as a gamification mechanism and as reliable cues of developers' ability. As the implementation of the feature appears to be in the initial stage, we will consider furthering this study, as changes to badges are hopefully introduced in GITHUB.

**Declaration of competing interest**

The authors declare the following financial interests/personal relationships which may be considered as potential competing interests: Fabio Calefato reports financial support was provided by The Italian Ministry for Universities and Research (MUR). Luigi Quaranta reports financial support was provided by The Italian Ministry for Universities and Research (MUR). Filippo Lanubile reports financial support was provided by The Italian Ministry for Universities and Research (MUR). If there are other authors, they declare that they have no known competing financial interests or personal relationships that could have appeared to influence the work reported in this paper.

**Data availability**

The manuscript contains a link to the full replication package hosted at zenodo.

**Acknowledgments**

This research was co-funded by projects DARE (PNC0000002, CUP: B53C22006420001), FAIR (PE00000013, CUP: H97G22000210007), SERICS (PE0000014, CUP: H93C22000620001), and QualAI (PRIN 2022, CUP: H53D23003510006). The authors thank the anonymous respondents to the survey and Prof. Claudia Capozza for her guidance and feedback on regression modeling.

## References

[1] J. Hamari, Gamification, in: The Blackwell Encyclopedia of Sociology, Wiley Online Library, 2007, pp. 1–3.
[2] D. Hunter, K. Werbach, For the Win, vol. 2, Wharton Digital Press, Philadelphia, PA, USA, 2012.
[3] S. Deterding, D. Dixon, R. Khaled, L.E. Nacke, From game design elements to gamefulness: defining ''gamification'', in: A. Lugmayr, H. Franssila, C. Safran, I. Hammouda (Eds.), Proc. 15th Int'l Academic MindTrek Conf.: Envisioning Future Media Environments, MindTrek 2011, Tampere, Finland, Sep. 28-30, 2011, ACM, 2011, pp. 9–15, http://dx.doi.org/10.1145/2181037.2181040.
[4] K. Robson, K. Plangger, J.H. Kietzmann, I. McCarthy, L. Pitt, Is it all a game? Understanding the principles of gamification, Bus. Horiz. 58 (4) (2015) 411–420.
[5] B. Vasilescu, V. Filkov, A. Serebrenik, StackOverflow and GitHub: Associations between software development and crowdsourced knowledge, in: Int'l Conf. on Social Computing, SocialCom 2013, Washington, DC, USA, 8-14 September 2013, IEEE Computer Society, 2013, pp. 188–195, http://dx.doi.org/10.1109/SocialCom.2013.35.
[6] S. Vessels, More contributions on your profile, 2016, https://github.blog/2016-05-19-more-contributions-on-your-profile/. (Accessed 19 August 2024).
[7] B. Brooks, Introducing achievements: Recognizing the many stages of a developer's coding journey, 2022, URL: https://github.blog/2022-06-09-introducing-achievements-recognizing-the-many-stages-of-a-developers-coding-journey. (Accessed 19 August 2024).
[8] L. Moldon, M. Strohmaier, J. Wachs, How gamification affects software developers: Cautionary evidence from a natural experiment on GitHub, in: 43rd IEEE/ACM Int'l Conf. on Software Engineering, ICSE 2021, Madrid, Spain, 22-30 May 2021, IEEE, 2021, pp. 549–561, http://dx.doi.org/10.1109/ICSE43902.2021.00058.
[9] C. Metz, How github conquered google, microsoft, and everyone else, 2015, URL: https://www.wired.com/2015/03/github-conquered-google-microsoft-everyone-else.
[10] Recruiting tech roles through GitHub, stack overflow, and open source contributions, 2022, URL: https://www.linkedin.com/pulse/recruiting-tech-roles-through-github-stack-overflow-openvitaetech/. (Accessed 19 August 2024).
[11] Discussion on hacker news: Recruiting tech roles through GitHub, stack overflow, and open source contributions, 2011, URL: https://news.ycombinator.com/item?id=2763182. (Accessed 19 August 2024).
[12] GitHub is your resume now, 2012, URL: https://anti-pattern.com/github-is-your-resume-now. (Accessed 19 August 2024).
[13] J. Marlow, L. Dabbish, Activity traces and signals in software developer recruitment and hiring, in: A.S. Bruckman, S. Counts, C. Lampe, L.G. Terveen (Eds.), Computer Supported Cooperative Work, CSCW 2013, San Antonio, TX, USA, February 23-27, 2013, ACM, 2013, pp. 145–156, http://dx.doi.org/10.1145/2441776.2441794.

[14] L. Singer, F.M.F. Filho, B. Cleary, C. Treude, M.D. Storey, K. Schneider, Mutual assessment in the social programmer ecosystem: an empirical investigation of developer profile aggregators, in: A.S. Bruckman, S. Counts, C. Lampe, L.G. Terveen (Eds.), Computer Supported Cooperative Work, CSCW 2013, San Antonio, TX, USA, February 23-27, 2013, ACM, 2013, pp. 103–116, http://dx.doi.org/10.1145/2441776.2441791.

[15] J. Tsay, L. Dabbish, J.D. Herbsleb, Influence of social and technical factors for evaluating contribution in GitHub, in: P. Jalote, L.C. Briand, A. van der Hoek (Eds.), 36th Int'l Conf. on Software Engineering, ICSE '14, Hyderabad, India - May 31 - June 07, 2014, ACM, 2014, pp. 356–366, http://dx.doi.org/10.1145/2568225.2568315.

[16] T. Erickson, W.A. Kellogg, Social translucence: an approach to designing systems that support social processes, ACM Trans. Comput. Hum. Interact. 7 (1) (2000) 59–83, http://dx.doi.org/10.1145/344949.345004.

[17] J.S. Donath, Signals in social supernets, J. Comput. Mediat. Commun. 13 (1) (2007) 231–251, http://dx.doi.org/10.1111/j.1083-6101.2007.00394.x.

[18] M. Spence, Job market signaling, in: Uncertainty in Economics, Elsevier, 1978, pp. 281–306.

[19] A. Zahavi, Mate selection—a selection for a handicap, J. Theor. Biol. 53 (1) (1975) 205–214.

[20] N.S. Shami, K. Ehrlich, G. Gay, J.T. Hancock, Making sense of strangers' expertise from signals in digital artifacts, in: D.R. Olsen Jr., R.B. Arthur, K. Hinckley, M.R. Morris, S.E. Hudson, S. Greenberg (Eds.), Proc. 27th Int'l Conf. on Human Factors in Computing Systems, CHI 2009, Boston, MA, USA, April 4-9, 2009, ACM, 2009, pp. 69–78, http://dx.doi.org/10.1145/1518701.1518713.

[21] J. Marlow, L. Dabbish, J.D. Herbsleb, Impression formation in online peer production: activity traces and personal profiles in github, in: A.S. Bruckman, S. Counts, C. Lampe, L.G. Terveen (Eds.), Computer Supported Cooperative Work, CSCW 2013, San Antonio, TX, USA, February 23-27, 2013, ACM, 2013, pp. 117–128, http://dx.doi.org/10.1145/2441776.2441792.

[22] L.A. Dabbish, H.C. Stuart, J. Tsay, J.D. Herbsleb, Social coding in GitHub: transparency and collaboration in an open software repository, in: S.E. Poltrock, C. Simone, J. Grudin, G. Mark, J. Riedl (Eds.), CSCW '12 Computer Supported Cooperative Work, Seattle, WA, USA, February 11-15, 2012, ACM, 2012, pp. 1277–1286, http://dx.doi.org/10.1145/2145204.2145396.

[23] J. Tsay, L. Dabbish, J.D. Herbsleb, Social media in transparent work environments, in: 6th Int'l Workshop on Cooperative and Human Aspects of Software Engineering, CHASE 2013, San Francisco, CA, USA, May 25, 2013, IEEE Computer Society, 2013, pp. 65–72, http://dx.doi.org/10.1109/CHASE.2013.6614733.

[24] D. de Paula Porto, G.M. de Jesus, F.C. Ferrari, S.C.P.F. Fabbri, Initiatives and challenges of using gamification in software engineering: A systematic mapping, J. Syst. Softw. 173 (2021) 110870, http://dx.doi.org/10.1016/j.jss.2020.110870.

[25] O. Pedreira, F. García, N.R. Brisaboa, M. Piattini, Gamification in software engineering - a systematic mapping, Inf. Softw. Technol. 57 (2015) 157–168, http://dx.doi.org/10.1016/J.INFSOF.2014.08.007.

[26] A. Anderson, D.P. Huttenlocher, J.M. Kleinberg, J. Leskovec, Steering user behavior with badges, in: D. Schwabe, V.A.F. Almeida, H. Glaser, R. Baeza-Yates, S.B. Moon (Eds.), 22nd International World Wide Web Conference, WWW '13, Rio de Janeiro, Brazil, May 13-17, 2013, International World Wide Web Conferences Steering Committee / ACM, 2013, pp. 95–106, http://dx.doi.org/10.1145/2488388.2488398.

[27] J. Fanfarelli, S. Vie, R. McDaniel, Understanding digital badges through feedback, reward, and narrative: A multidisciplinary approach to building better badges in social environments, Commun. Des. Q. Rev. 3 (3) (2015) 56–60, http://dx.doi.org/10.1145/2792989.2792998.

[28] M. Papoutsoglou, G.M. Kapitsaki, L. Angelis, Modeling the effect of the badges gamification mechanism on personality traits of stack overflow users, Simul. Model. Pract. Theory 105 (2020) 102157, http://dx.doi.org/10.1016/J.SIMPAT.2020.102157.

[29] A. Trockman, S. Zhou, C. Kästner, B. Vasilescu, Adding sparkle to social coding: an empirical study of repository badges in the *npm* ecosystem, in: M. Chaudron, I. Crnkovic, M. Chechik, M. Harman (Eds.), Proc. 40th Int'l Conf. on Software Engineering, ICSE 2018, Gothenburg, Sweden, May 27 - June 03, 2018, ACM, 2018, pp. 511–522, http://dx.doi.org/10.1145/3180155.3180209.

[30] D. Legay, A. Decan, T. Mens, On the usage of badges in open source packages on GitHub, in: D.D. Nucci, C.D. Roover (Eds.), Proc. 18th Belgium-Netherlands Software Evolution Workshop, Brussels, Belgium, November 28th To 29th, 2019, in: CEUR Workshop Proceedings, vol. 2605, CEUR-WS.org, 2019, URL: https://ceur-ws.org/Vol-2605/9.pdf.

[31] A.S.M. Venigalla, K. Boyalakuntla, S. Chimalakonda, GitQ - towards using badges as visual cues for GitHub projects, in: A. Rastogi, R. Tufano, G. Bavota, V. Arnaoudova, S. Haiduc (Eds.), Proc. 30th IEEE/ACM Int'L Conf. on Program Comprehension, ICPC 2022, May 16-17, 2022, ACM, 2022, pp. 157–161, http://dx.doi.org/10.1145/3524610.3527876.

[32] W. Hammedi, T. Leclercq, I. Poncin, L. Alkire (Née Nasr), Uncovering the dark side of gamification at work: Impacts on engagement and well-being, J. Bus. Res. (ISSN: 0148-2963) 122 (2021) 256–269, http://dx.doi.org/10.1016/j.jbusres.2020.08.032, URL: https://www.sciencedirect.com/science/article/pii/S0148296320305415.

[33] F. Calefato, F. Lanubile, N. Novielli, How to ask for technical help? Evidence-based guidelines for writing questions on stack overflow, Inf. Softw. Technol. 94 (2018) 186–207, http://dx.doi.org/10.1016/j.infsof.2017.10.009.

[34] S. Grant, B. Betts, Encouraging user behaviour with achievements: an empirical study, in: T. Zimmermann, M.D. Penta, S. Kim (Eds.), Proc. 10th Working Conf. on Mining Software Repositories, MSR '13, San Francisco, CA, USA, May 18-19, 2013, IEEE Computer Society, 2013, pp. 65–68, http://dx.doi.org/10.1109/MSR.2013.6624007.

[35] F. Calefato, M.A. Gerosa, G. Iaffaldano, F. Lanubile, I. Steinmacher, Will you come back to contribute? Investigating the inactivity of OSS core developers in GitHub, Empir. Softw. Eng. 27 (3) (2022) 76, http://dx.doi.org/10.1007/s10664-021-10012-6.

[36] E. Kalliamvakou, G. Gousios, K. Blincoe, L. Singer, D.M. Germán, D.E. Damian, The promises and perils of mining GitHub, in: P.T. Devanbu, S. Kim, M. Pinzger (Eds.), 11th Working Conference on Mining Software Repositories, MSR 2014, Proceedings, May 31 - June 1, 2014, Hyderabad, India, ACM, 2014, pp. 92–101, http://dx.doi.org/10.1145/2597073.2597074.

[37] Y. Zhang, M. Qin, K. Stol, M. Zhou, H. Liu, How are paid and volunteer open source developers different? A study of the rust project, in: Proc. of the 46th IEEE/ACM Int'l Conf. on Software Engineering, ICSE 2024, Lisbon, Portugal, April 14-20, 2024, ACM, 2024, pp. 195:1–195:13, http://dx.doi.org/10.1145/3597503.3639197.

[38] J.D. Angrist, J.-S. Pischke, Mostly Harmless Econometrics: An Empiricist's Companion, Princeton University Press, 2009.

[39] M. Lechner, The Estimation of Causal Effects by Difference-in-Difference Methods, in: Foundations and Trends in Econometrics, NEW PUBL INC, ISBN: 1601984987, 2024, p. 72.

[40] J. Romano, J.D. Kromrey, J. Coraggio, J. Skowronek, Appropriate statistics for ordinal level data: Should we really be using t-test and Cohen'sd for evaluating group differences on the NSSE and other surveys, in: Annual Meeting of the Florida Association of Institutional Research, Vol. 177, 2006, p. 34.

[41] G. Gousios, A. Zaidman, M.D. Storey, A. van Deursen, Work practices and challenges in pull-based development: The integrator's perspective, in: A. Bertolino, G. Canfora, S.G. Elbaum (Eds.), 37th IEEE/ACM Int'l Conf. on Software Engineering, ICSE 2015, Florence, Italy, May 16-24, 2015, Volume 1, IEEE Computer Society, 2015, pp. 358–368, http://dx.doi.org/10.1109/ICSE.2015.55.

[42] G. Gousios, M.D. Storey, A. Bacchelli, Work practices and challenges in pull-based development: the contributor's perspective, in: L.K. Dillon, W. Visser, L.A. Williams (Eds.), Proc. 38th Int'l Conf. on Software Engineering, ICSE 2016, Austin, TX, USA, May 14-22, 2016, ACM, 2016, pp. 285–296, http://dx.doi.org/10.1145/2884781.2884826.

[43] J.L.C. Izquierdo, J. Cabot, On the analysis of non-coding roles in open source development, Empir. Softw. Eng. 27 (1) (2022) 18, http://dx.doi.org/10.1007/s10664-021-10061-x.

[44] K. Blincoe, J. Sheoran, S.P. Goggins, E. Petakovic, D.E. Damian, Understanding the popular users: Following, affiliation influence and leadership on GitHub, Inf. Softw. Technol. 70 (2016) 30–39, http://dx.doi.org/10.1016/j.infsof.2015.10.002.

[45] H. Hata, N. Novielli, S. Baltes, R.G. Kula, C. Treude, GitHub discussions: An exploratory study of early adoption, Empir. Softw. Eng. 27 (1) (2022) 3, http://dx.doi.org/10.1007/s10664-021-10058-6.

[46] Y. Lu, X. Mao, M. Zhou, Y. Zhang, Z. Li, T. Wang, G. Yin, H. Wang, Motivation under gamification: An empirical study of developers' motivations and contributions in stack overflow, IEEE Trans. Softw. Eng. 48 (12) (2022) 4947–4963, http://dx.doi.org/10.1109/TSE.2021.3130088.

[47] A. Bosu, C.S. Corley, D. Heaton, D. Chatterji, J.C. Carver, N.A. Kraft, Building reputation in StackOverflow: an empirical investigation, in: T. Zimmermann, M.D. Penta, S. Kim (Eds.), Proc. of 10th Working Conference on Mining Software Repositories, MSR '13, San Francisco, CA, USA, May 18-19, 2013, IEEE Computer Society, 2013, pp. 89–92, http://dx.doi.org/10.1109/MSR.2013.6624013.

[48] A.C. Cameron, P.K. Trivedi, Microeconometrics: Methods and Applications, Cambridge University Press, 2005.

[49] T.D. Sasso, A. Mocci, M. Lanza, E. Mastrodicasa, How to gamify software engineering, in: M. Pinzger, G. Bavota, A. Marcus (Eds.), IEEE 24th Int'l Conf. on Software Analysis, Evolution and Reengineering, SANER 2017, Klagenfurt, Austria, February 20-24, 2017, IEEE Computer Society, 2017, pp. 261–271, http://dx.doi.org/10.1109/SANER.2017.7884627.

[50] F. García, O. Pedreira, M. Piattini, A. Cerdeira-Pena, M.R. Penabad, A framework for gamification in software engineering, J. Syst. Softw. 132 (2017) 21–40, http://dx.doi.org/10.1016/j.jss.2017.06.021.

[51] T.D. Sasso, A. Mocci, M. Lanza, E. Mastrodicasa, How to gamify software engineering, in: M. Pinzger, G. Bavota, A. Marcus (Eds.), IEEE 24th Int'l Conf. on Software Analysis, Evolution and Reengineering, SANER 2017, Klagenfurt, Austria, February 20-24, 2017, IEEE Computer Society, 2017, pp. 261–271, http://dx.doi.org/10.1109/SANER.2017.7884627.

[52] B. Morschheuser, L. Hassan, K. Werder, J. Hamari, How to design gamification? A method for engineering gamified software, Inf. Softw. Technol. 95 (2018) 219–237, http://dx.doi.org/10.1016/j.infsof.2017.10.015.

[53] 70% Of employers are snooping candidates' social media, 2017, URL: https://www.careerbuilder.com/advice/blog/social-media-survey-2017. (Accessed 19 August 2024).

[54] D. Bohnert, W.H. Ross, The influence of social networking web sites on the evaluation of job candidates, Cyberpsychol. Behav. Soc. Netw. 13 (3) (2010) 341–347, http://dx.doi.org/10.1089/CYBER.2009.0193.

[55] Y. Lu, X. Mao, M. Zhou, Y. Zhang, Z. Li, T. Wang, G. Yin, H. Wang, Motivation under gamification: An empirical study of developers' motivations and contributions in stack overflow, IEEE Trans. Softw. Eng. 48 (12) (2022) 4947–4963, http://dx.doi.org/10.1109/TSE.2021.3130088.

[56] A. Sarma, X. Chen, S.K. Kuttal, L. Dabbish, Z. Wang, Hiring in the global stage: Profiles of online contributions, in: 11th IEEE Int'l Conf. on Global Software Engineering, ICGSE 2016, Orange County, CA, USA, August 2-5, 2016, IEEE Computer Society, 2016, pp. 1–10, http://dx.doi.org/10.1109/ICGSE.2016.35.

[57] A. Capiluppi, A. Serebrenik, L. Singer, Assessing technical candidates on the social web, IEEE Softw. 30 (1) (2013) 45–51, http://dx.doi.org/10.1109/MS.2012.169.

[58] G.J. Greene, B. Fischer, Cvexplorer: identifying candidate developers by mining and exploring their open source contributions, in: D. Lo, S. Apel, S. Khurshid (Eds.), Proc. 31st IEEE/ACM Int'l Conf. on Automated Software Engineering, ASE 2016, Singapore, September 3-7, 2016, ACM, 2016, pp. 804–809, http://dx.doi.org/10.1145/2970276.2970285.

[59] J.B. Walther, B.V.D. Heide, L.M. Hamel, H.C. Shulman, Self-generated versus other-generated statements and impressions in computer-mediated communication, Commun. Res. 36 (2) (2009) 229–253, http://dx.doi.org/10.1177/0093650208330251.